# Analysis of Eccentric Coaxial Waveguides Filled with Lossy Anisotropic Media via Finite Difference

Raul O. Ribeiro and Maria A. Martinez
Department of Electrical Engineering
Federal Center for Technological Education
Celso Suckow da Fonseca (CEFET/RJ)
Rio de Janeiro, RJ, Brazil

Guilherme S. Rosa and Rafael A. Penchel
Department of Electrical Engineering
São Paulo State University (UNESP)
São João da Boa Vista, SP, Brazil

***Abstract*—This study presents a finite difference method (FDM) to model the electromagnetic field propagation in eccentric coaxial waveguides filled with lossy uniaxially anisotropic media. The formulation utilizes conformal transformation to map the eccentric circular waveguide into an equivalent concentric one. In the concentric problem, we introduce a novel normalized Helmholtz equation to decouple TM and TE modes, and we solve this non-homogeneous partial differential equation using the finite difference in cylindrical coordinates. The proposed approach was validated against perturbation-based, spectral element-based, and finite-integration-based numerical solutions. The preliminary results show that our solution is superior in computational time. Furthermore, our FDM formulation can be extended with minimal adaptations to model complex media problems, such as metamaterial devices, optical fibers, and geophysical exploration sensors.**

## I. Introduction

Conformal mapping is a robust technique for solving a range of electromagnetic problems. Its primary concept involves mapping complex geometries into more tractable domains via analytical transformations. Within the mapped domain, both analytical and numerical techniques can be effectively applied. In [1], TE and TM modes were calculated in transmission lines with an eccentric circular inner conductor. In [2], conformal mapping was used to transform an eccentric coaxial waveguide into an equivalent concentric problem, where the cutoff wavenumbers for TM and TE modes were found from the solution of the weighted Helmholtz equation. The concentric problem was solved by finite difference in [3]. In [4], the eccentric domain was mapped onto the rectangular domain, and the wavenumbers were obtained via the finite element method (FEM). In [5], transformation optics (TO) was employed to map the eccentric problem into a concentric problem, and the cutoff wavenumbers and field patterns were obtained using perturbation techniques. Similarly, in [6], [7], the eccentric domain was mapped onto the concentric domain via TO, and the mapped problem was solved using the spectral element method (SEM).

The present work employs a conformal transformation to map the eccentric coaxial waveguide into an equivalent concentric waveguide. In the concentric domain, we introduce a novel normalized Helmholtz equation to decouple TM and TE modes. We solve this non-homogeneous partial differential equation using the finite difference method (FDM) in cylindrical coordinates. In contrast to [3], our formulation can model waveguides filled with anisotropic media. Unlike perturbation-based techniques [5], our approach does not impose restrictions on small eccentricities. Moreover, the proposed normalized formulation can obtain an entire set of solutions independently of the medium's parameters, based solely on the problem's geometry.

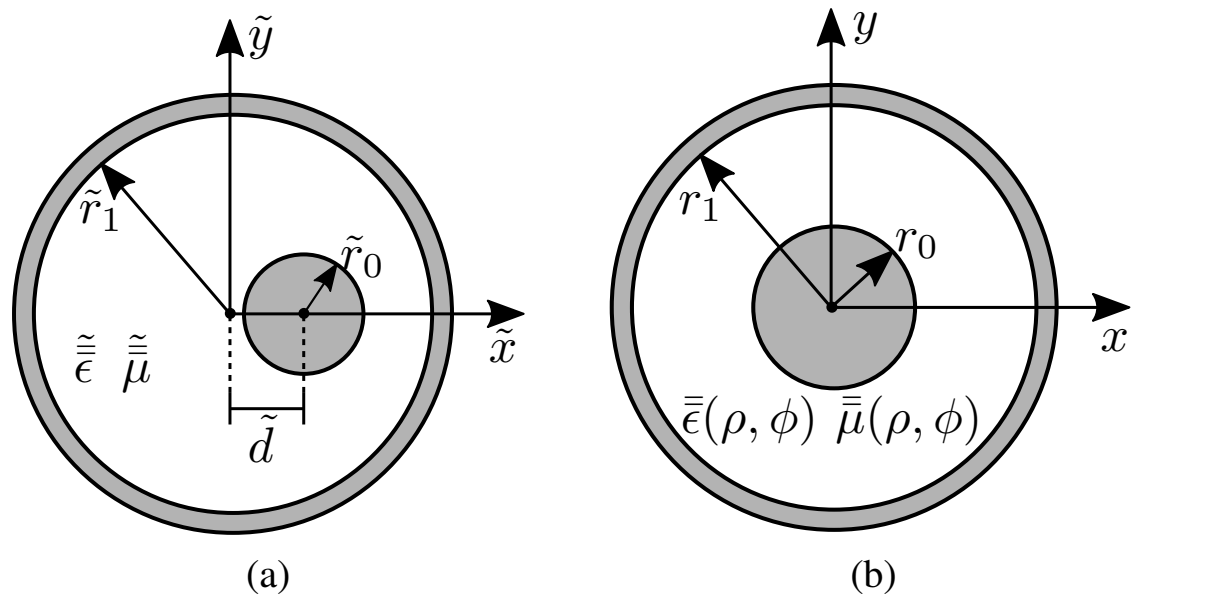

Fig. 1: (a) Cross-section $\tilde{\Omega}$ of an eccentric coaxial waveguide invariant along the axial direction. (b) Cross-section $\Omega$ of the mapped concentric coaxial waveguide.

## II. Formulation Overview

### A. *Normalized Helmholtz Equation*

Consider a non-concentric coaxial waveguide that remains constant along the axial direction and confined by a perfect electric conductor (PEC) with a radius of $\tilde{r}_1$. The inner conductor is a circular PEC cylinder with a radius of $\tilde{r}_0$ and positioned at a distance $\tilde{d}$ from the $\tilde{z}$-axis. This eccentric structure is described in a cylindrical coordinate system denoted by $(\tilde{\rho}, \tilde{\phi}, \tilde{z})$, with the corresponding cross-section $\tilde{\Omega}$ shown in Fig. 1a. We assume a time-harmonic dependence of the form $e^{-i\omega t}$ and that the waveguide is filled with lossy uniaxially anisotropic media characterized by real-valued permeability tensor[1]

$$\tilde{\bar{\bar{\mu}}} = \mu_0 \tilde{\bar{\bar{\mu}}}_r \text{ with } \tilde{\bar{\bar{\mu}}}_r = \operatorname{diag}(\tilde{\mu}_{rs}, \tilde{\mu}_{rs}, \tilde{\mu}_{rz}), \tag{1}$$

[1]Complex permeability can also be considered by allowing $\tilde{\mu}_{rs}$ and $\tilde{\mu}_{rz}$ to be complex-valued parameters.

and the complex-valued permittivity tensor

$$\bar{\bar{\epsilon}} = \epsilon_0 \bar{\bar{\epsilon}}_r + \frac{i}{\omega}\bar{\bar{\sigma}}, \tag{2}$$

with

$$\bar{\bar{\epsilon}}_r = \text{diag}(\tilde{\epsilon}_{rs}, \tilde{\epsilon}_{rs}, \tilde{\epsilon}_{rz}),\ \bar{\bar{\sigma}} = \text{diag}(\tilde{\sigma}_s, \tilde{\sigma}_s, \tilde{\sigma}_z), \tag{3}$$

where $\bar{\bar{\sigma}}$ represents the conductive tensor, and $\omega = 2\pi f$ is angular frequency.

From conformal transformation optics [8]–[10], the original electromagnetic problem in the eccentric coordinates $(\tilde{\rho}, \tilde{\phi}, \tilde{z})$ can be mapped into a concentric problem with coordinates $(\rho, \phi, z)$ using

$$\mathbf{F} = \bar{\bar{J}}_{\text{TO}} \cdot \tilde{\mathbf{F}}, \text{ with } \mathbf{F} \in \{\mathbf{E}, \mathbf{H}\}, \tag{4}$$

$$\bar{\bar{p}} = |\bar{\bar{J}}_{\text{TO}}|^{-1} \bar{\bar{J}}_{\text{TO}} \cdot \bar{\bar{\tilde{p}}} \cdot \bar{\bar{J}}_{\text{TO}}^T, \text{ with } p \in \{\mu, \epsilon\}, \tag{5}$$

where $\bar{\bar{J}}_{\text{TO}}$ is the Jacobian of the transformation $(\tilde{\rho}, \tilde{\phi}, \tilde{z}) \to (\rho, \phi, z)$, and $|\bar{\bar{J}}_{\text{TO}}|$ is the respectively determinant. Assuming $z = \tilde{z}$ and $r_1 = \tilde{r}_1$, and uniaxially anisotropic media, we can simplify (5) and express the transformed constitutive tensor as

$$\bar{\bar{p}} = \text{diag}(p_s, p_s, p_z(\rho, \phi)) = \text{diag}(\tilde{p}_s, \tilde{p}_s, |\bar{\bar{J}}_{\text{TO}}|^{-1}\tilde{p}_z), \tag{6}$$

with $p \in \{\mu, \epsilon\}$, where

$$|\bar{\bar{J}}_{\text{TO}}|^{-1} = \frac{(1 - \tilde{x}_1/\tilde{x}_2)^2}{(1 - 2\rho\cos\phi/\tilde{x}_2 + \rho^2/\tilde{x}_2^2)^2}, \tag{7}$$

and

$$\tilde{x}_{1,2} = \frac{-\tilde{c} \mp \sqrt{\tilde{c}^2 - 4\tilde{r}_1^2}}{2}, \text{ with } \tilde{c} = \frac{\tilde{r}_0^2 - \tilde{r}_1^2 - \tilde{d}^2}{\tilde{d}}. \tag{8}$$

In the transformed domain, the axial constitutive parameters $\mu_z$ and $\epsilon_z$ change with respect to both $\rho$ and $\phi$ directions. This differs from the assumption of constant tensors in (1) and (2).

The problem in $\Omega$-domain (depicted in Fig. 1b) can be solved by decomposing electromagnetic fields into a sum of transverse magnetic (TM) and transverse electric (TE) modes [11], [12]. It is convenient to solve $E_z$ and $H_z$ axial field components and then compute the transverse components of the fields in cylindrical coordinates. The scalar Helmholtz equation for $E_z$ and $H_z$ in the concentric domain satisfies

$$\left(\nabla_s^2 + \frac{p_z(\rho, \phi)}{p_s} k_\rho^2\right) F = 0, \tag{9}$$

where $\nabla_s^2$ is the transverse Laplacian operator, $k_\rho$ is the radial wavenumber, $F = E_z$ if $p_{s,z} = \epsilon_{s,z}$, and $F = H_z$ if $p_{s,z} = \mu_{s,z}$. The domain is truncated by a PEC; then the fields must satisfy $F = 0$ for $F = E_z$ and $\partial F/\partial\rho = 0$ for $F = H_z$ on $\rho \in \{r_0, r_1\}$.

From (6), using $p_s = \tilde{p}_s$ and $p_z = \tilde{p}_z|\bar{\bar{J}}_{\text{TO}}|^{-1}$, we can rewrite (9) and define a normalized Helmholtz equation as follows

$$\nabla_s F = \lambda\ |\bar{\bar{J}}_{\text{TO}}(\rho, \phi)|^{-1} F, \tag{10}$$

where $\lambda = -(\tilde{p}_z/\tilde{p}_s)k_\rho^2$. For uniaxially anisotropic media, it's crucial to note that the axial field distribution in the $\Omega$-domain is solely determined by the problem's geometry. Therefore, once the values of $\tilde{r}_1$, $\tilde{r}_0$, and $\tilde{d}$ are defined, an entire set of solutions is obtained that is independent of the medium's parameters.

### *B. Finite Difference Method*

As a result of the conformal transformation [10], the normalized Helmholtz equation in the concentric domain (10) can be written in cylindrical coordinates as

$$\frac{1}{\rho}\frac{\partial F}{\partial \rho} + \frac{\partial^2 F}{\partial \rho^2} + \frac{1}{\rho^2}\frac{\partial^2 F}{\partial \phi^2} = \lambda\ |\bar{\bar{J}}_{\text{TO}}(\rho, \phi)|^{-1} F. \tag{11}$$

This is a non-homogeneous partial differential equation, which we address using the finite difference method [13], [14]. Consider the coaxial domain discretized into $M$ nodes in the $\rho$-direction and $N$ nodes in the $\phi$-direction. The grid can be defined by $\rho_i = r_0 + (i-1)h_\rho$ for $i = 1, 2, ..., M$, and $\phi_j = (j-1)h_\phi$ for $j = 1, 2, ..., N$, where

$$h_\rho = \frac{r_1 - r_0}{M - 1} \text{ and } h_\phi = \frac{2\pi}{N - 1}. \tag{12}$$

We approximate the derivatives in (11) using second-order central difference

$$\frac{1}{\rho_i}\frac{\partial F}{\partial \rho} \approx \frac{F(\rho_i + h_\rho, \phi_j) - F(\rho_i - h_\rho, \phi_j)}{2\rho_i h_\rho} \tag{13}$$

$$\frac{\partial^2 F}{\partial \rho^2} \approx \frac{F(\rho_i - h_\rho, \phi_j) - 2F(\rho_i, \phi_j) + F(\rho_i + h_\rho, \phi_j)}{h_\rho^2} \tag{14}$$

$$\frac{1}{\rho_i^2}\frac{\partial^2 F}{\partial \phi^2} \approx \frac{F(\rho_i, \phi_j - h_\phi) - 2F(\rho_i, \phi_j) + F(\rho_i, \phi_j + h_\phi)}{\rho_i^2 h_\phi^2}. \tag{15}$$

Additionally, we need to satisfy the $2\pi$-periodic condition

$$F(\rho_i, \phi_1) = F(\rho_i, \phi_N)\ \ \forall i = 1, 2, ..., M, \tag{16}$$

and the PEC condition for the TM case

$$F(\rho_i, \phi_j) = 0 \text{ for } i = 1, M,\ \forall j = 1, 2, ..., N, \tag{17}$$

and the TE case

$$F(\rho_0, \phi_j) = F(\rho_2, \phi_j) \text{ and } F(\rho_{M-1}, \phi_j) = F(\rho_{M+1}, \phi_j) \tag{18}$$

for all $j = 1, 2, ..., N$.

Using the finite difference (13)-(15) in the normalized wave equation (11), we can define the generalized eigenvalue problem

$$\mathbf{A}\bar{v} = \lambda \mathbf{B}\bar{v}, \tag{19}$$

where $\lambda$ and $\bar{v}$ represent eigenvalues and eigenvectors, respectively. The eigenvalues are associated with the radial wavenumbers via

$$k_\rho = \sqrt{-\frac{\tilde{p}_s}{\tilde{p}_z}\lambda}, \tag{20}$$

where $p_{s,z} = \epsilon_{s,z}$ for TM modes and $p_{s,z} = \mu_{s,z}$ for TE modes, while the eigenvectors are associated with the axial field components. The axial wavenumbers can be obtained in the form

$$k_z = \sqrt{k_s^2 - k_\rho^2}. \tag{21}$$

The linear system (19) has $MN$ degrees of freedom for the TE modes. In the TM case, the PEC boundary condition $E_z = 0$ must be satisfied for $\rho \in \{r_0, r_1\}$, which implies in $(M-2)N$ DoF.

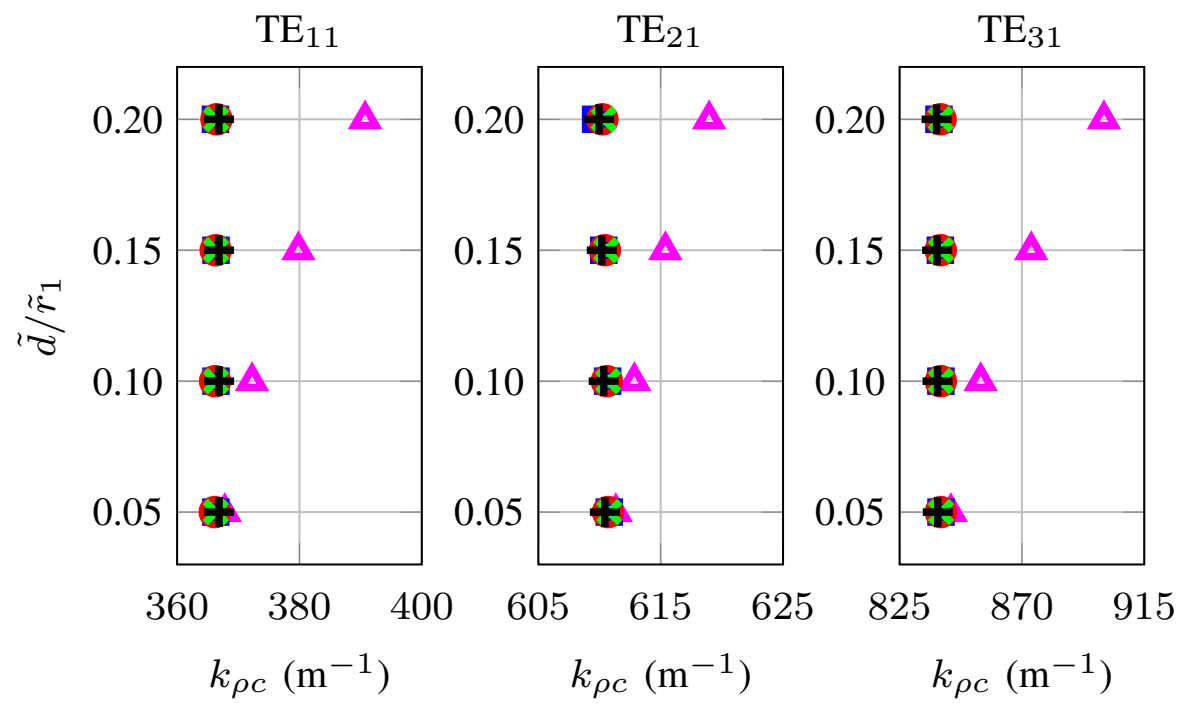

(a)

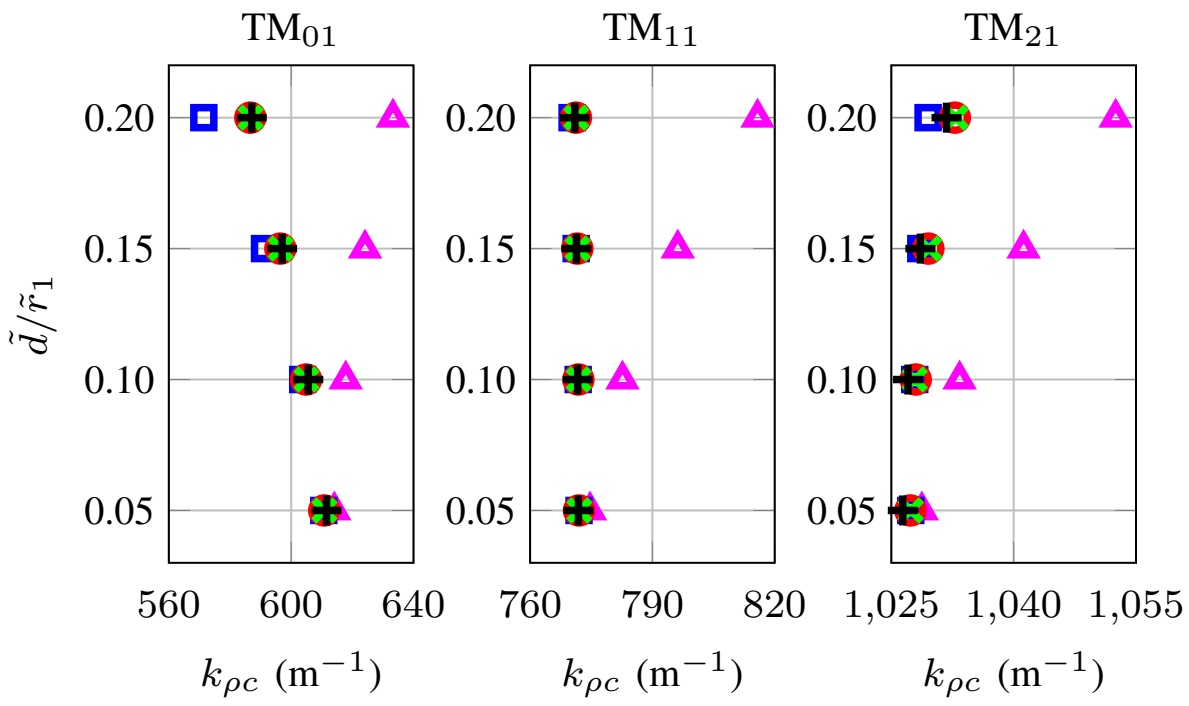

(b)

Fig. 2: Radial wavenumbers for (a) TE and (b) TM modes as a function of the normalized distance $\tilde{d}/\tilde{r}_1$ get by CMPM (△), by RPM (□), by FEM (○), by SEM (×), and by FDM (+).

## III. Numerical Results

For our initial validation, we examine an eccentric waveguides characterized by $\tilde{r}_1 = 5$ mm, $\tilde{r}_0 = 0.05\tilde{r}_1$, and different eccentricity offsets $\tilde{d} \in \{0.05\tilde{r}_1, 0.10\tilde{r}_1, 0.15\tilde{r}_1, 0.20\tilde{r}_1\}$, while assuming the medium is the vacuum. This problem was previously studied in [5]–[7] and is selected here to investigate the accuracy of the present finite difference method. Fig. 2 shows the radial wavenumber of the first three TE and TM modes as a function of the normalized offset $\tilde{d}/\tilde{r}_1$ using the cavity-material perturbation method (CMPM) [15], the regular perturbation method (RPM) from [5], FEM from [16], SEM present in [6], [7], and our FDM formulation. Notice that CMPM does not yield accurate results for $\tilde{d} \geq 0.1\tilde{r}_1$. In particular, the $TM_{01}$ and $TM_{21}$ results from RPM present a noticeable error for $\tilde{d} = 0.2\tilde{r}_1$. The increase in the relative error with the rise in the normalized displacement $\tilde{d}/\tilde{r}_1$ of the inner conductor poses a constraint on perturbation methods. In contrast, the FDM exhibited strong agreement across all simulated scenarios compared to FEM and SEM solutions. This suggests that the FDM approach can be applied to geometries with more substantial deviations of the inner conductor.

With the second example, we explore the capacity of our finite difference formulation to address eccentric waveguides filled with lossless uniaxial anisotropic media. For this, we reproduce the cases considered in [5]–[7], in which $\tilde{\epsilon}_s = \epsilon_0 \tilde{\epsilon}_{rs}$, $\tilde{\epsilon}_z = \epsilon_0 \tilde{\epsilon}_{rz}$, and $\tilde{\mu}_s = \tilde{\mu}_z = \mu_0$, where $\epsilon_0$ and $\mu_0$ are the vacuum values. The geometry has $\tilde{r}_1 = 5$ mm, $\tilde{r}_0 = 0.05\tilde{r}_1$, and offset $\tilde{d} = 0.2\tilde{r}_1$. Figs. 3 and 4 show results for the radial wavenumbers of the $TM_{01}$ mode as a function of $\tilde{\epsilon}_{rs}$ and $\tilde{\epsilon}_{rz}$. We can observe excellent agreement between dense-mesh finite-integration technique (FIT) [16], SEM, and FDM. However, the CMPM and RPM present a visible error. One significant benefit of our normalized approach is that we only need to compute once the eigenvalues related to the problem's geometry. When there are changes to the media parameters, we can recompute the radial wavenumbers using (20).

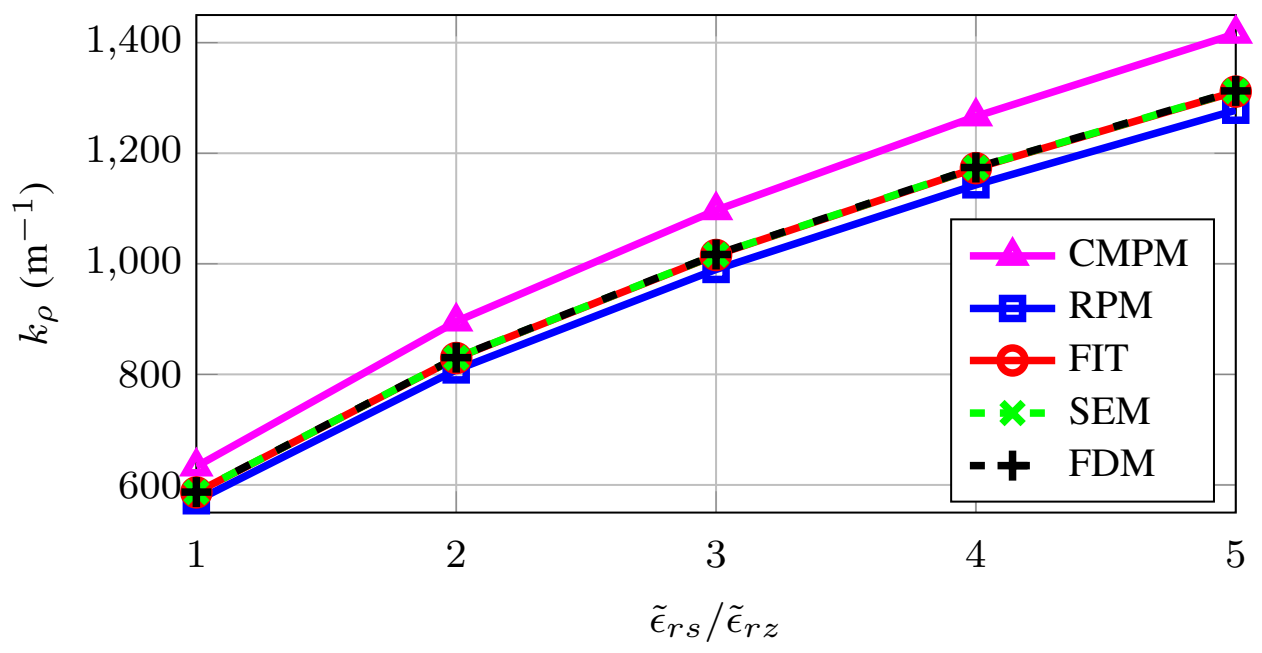

Fig. 3: Radial wavenumbers for the $TM_{01}$ mode as a function of the anisotropic permittivity ratio $\tilde{\epsilon}_{rs}/\tilde{\epsilon}_{rz}$, with $\tilde{\epsilon}_{rz} = 1$.

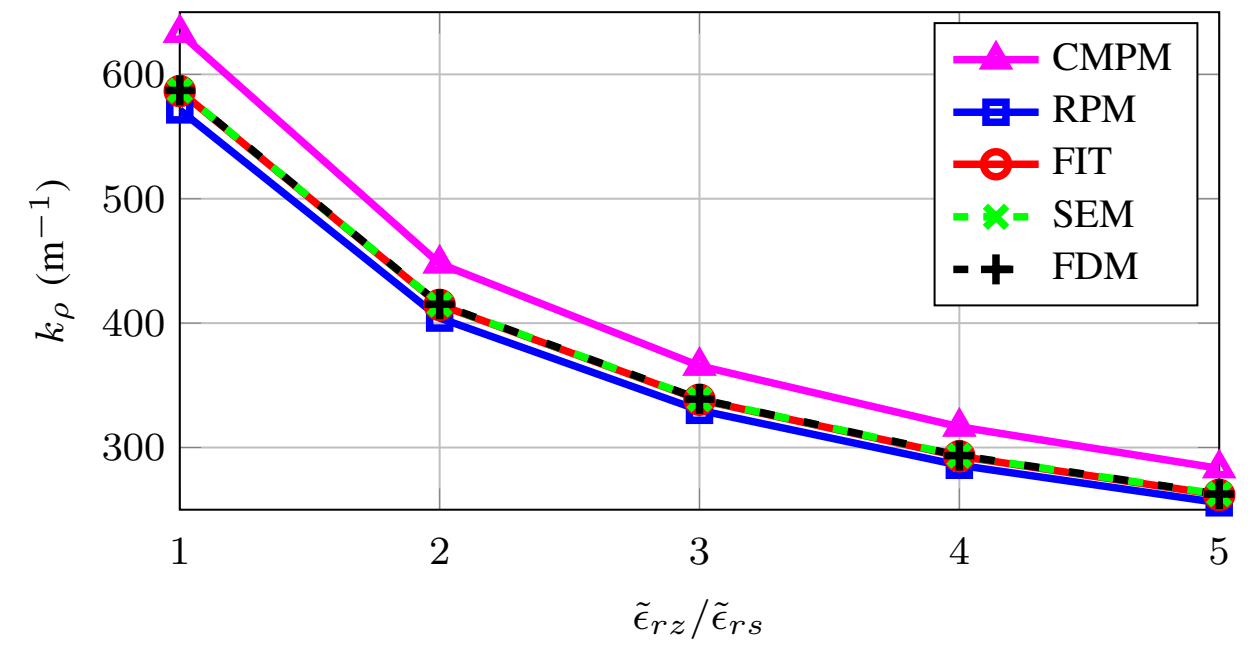

Fig. 4: Radial wavenumbers for the $TM_{01}$ mode as a function of the anisotropic permittivity ratio $\tilde{\epsilon}_{rz}/\tilde{\epsilon}_{rs}$, with $\tilde{\epsilon}_{rs} = 1$.

The computational time cost required by the FDM (considering $M = 10$ and $N = 41$, which implies 328 DoFs) is compared with that of the FIT, CMPM, RPM, and SEM in Table I for the anisotropic waveguide with $\tilde{\epsilon}_{rs} = 5$ and $\tilde{\epsilon}_{rz} = 1$ (Case I), with $\tilde{\epsilon}_{rz} = 5$ and $\tilde{\epsilon}_{rs} = 1$ (Case II). The present FDM has a computation time similar to the SEM and significantly lowers costs compared to the other techniques. The FDM and SEM algorithms were written in Matlab [17] running on a PC with 1.80 GHz Intel Core i7-8565U with four cores. The FIT, CMPM, and RMP times were obtained from [5].

Lastly, we examine the capability of our FDM formulation to solve eccentric waveguides filled with lossy media. The media under investigation exhibit uniaxial anisotropy, as de-scribed by the real permittivity tensor $\tilde{\bar{\bar{\epsilon}}}_r = (5.6, 5.6, 4.6)$ and the conductive tensor $\tilde{\bar{\bar{\sigma}}} = (0.38, 0.38, 0.34)$ to characterize the media dissipation. The dimensions of the waveguide are defined as $\tilde{r}_1 = 10$ mm, $\tilde{r}_0 = 0.2\tilde{r}_1$, and $\tilde{d} = 0.3\tilde{r}_1$. We present the results for the real and imaginary components of the axial wavenumber $k_z$ as a function of the operating frequency $f$, ranging from 1 GHz to 10 GHz, in Fig. 5. For this example, we consider the same mesh as before. The FDM results are in excellent agreement with the FIT results.

TABLE I: Computational cost.

| | FIT | CMPM | RPM | SEM | FDM |
|---|---|---|---|---|---|
| Case I | 15 min 44 s | 2.07 s | 16.72 s | 0.64 s | 0.15 s |
| Case II | 22 min 27.10 s | 2.43 s | 21.60 s | 0.63 s | 0.15 s |

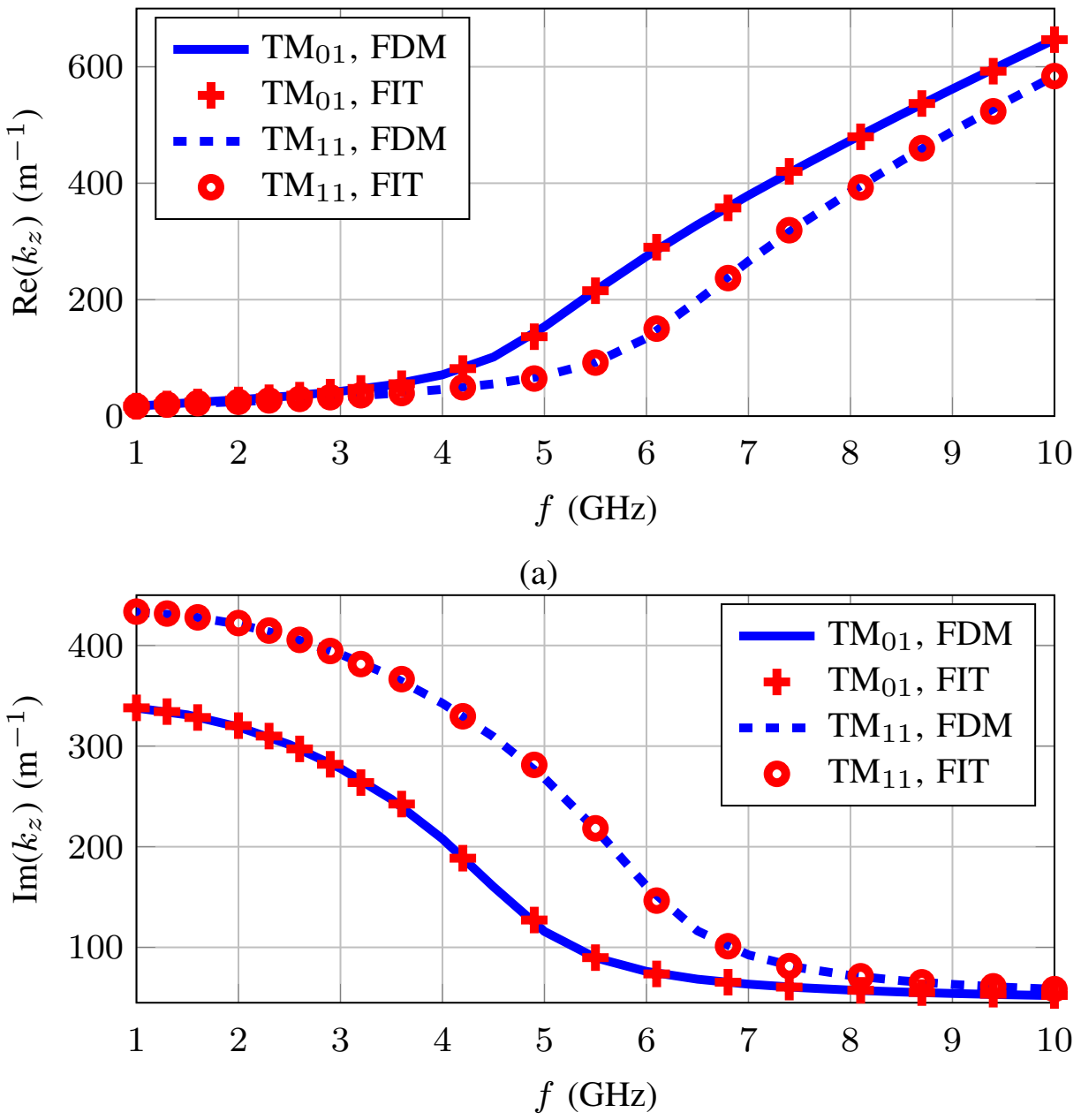

(a)

(b)

Fig. 5: Axial wavenumbers for the $\mathrm{TM}_{01}$ and $\mathrm{TM}_{11}$ modes as a function of frequency.

It's important to mention that according to (2), frequency variation affects the imaginary part of the complex permittivity, leading to changes in the medium filling the waveguide. Since the geometric parameters remain unchanged, we only need to compute the eigenvalues of this structure once. Afterward, the radial wavenumbers $k_\rho$ are determined using (20), and the axial wavenumbers $k_z$ are calculated via (21).

## IV. Conclusion

This work explored a finite difference formulation to calculate the radial and axial wavenumbers in eccentric coaxial waveguides filled with lossy uniaxially anisotropic media. Given the problem's geometry, the proposed normalized wave equation obtains an entire set of solutions independently of the medium's parameters. The method utilizes conformal transformation to map the original eccentric waveguides into equivalent concentric problems. In contrast to perturbation-based techniques, our approach does not limit the eccentricity offset and offers excellent accuracy with minimal computational cost. Our future research aims to develop models for concentric and non-concentric multilayered cylindrical waveguides filled with anisotropic media. This is crucial for various applications, including high-frequency optical circuits, multi-core fibers, and low-frequency electromagnetic sensors, which predict oil presence in geophysical formations.

## Acknowledgment

This work was partly supported by the Brazilian agency CAPES under Grant 88887.974602/2024-00.